# Advancing Evidence Generation in Biomedical Research Using Natural Hermite and Propensity Score Indices: Applications to External Control Arms

J. Cabrera, B. Alemayehu, D. Alemayehu, and S. Weigle

**Abstract**

When it is not feasible to conduct randomized controlled trials (RCTs), the use of external control arms based on real-world data (RWD) may be a viable option. However, challenges arising from data heterogeneity must be addressed to ensure the reliability of trial results. We consider the use of Natural Hermite and propensity score indices to facilitate robust comparisons between RCTs and RWD studies. Illustrations are provided on the implementation and performance of the underlying algorithms using simulated data, as well as synthetic data from a clinical trial and RWD.

1. Introduction

Traditional randomized controlled trials (RCTs)[9][10] typically employ a control arm comprising subjects who receive either standard treatment or a placebo. While RCTs are considered the gold standard for generating reliable evidence, recruiting patients for the control arm can be challenging in certain situations, especially in rare disease research, due to operational constraints or, in the case of placebo, ethical concerns. Often, the effect of the standard treatment is well established because of historical clinical trials conducted to evaluate it. Furthermore, large real-world datasets—such as claims data or electronic health records (EHRs)—exist and contain historical information from many patients who have received the standard treatment.

In recent years, the use of external control arms (ECAs) to augment RCTs has attracted considerable interest among researchers conducting clinical trials. Nevertheless, there are significant methodological and practical challenges that must be addressed for the successful implementation of ECAs. One of the primary concerns is ensuring the comparability of patients in external control and trial arms. Various approaches have been proposed in the literature [1] to address this issue, including Bayesian borrowing methods [2], [3] and propensity score (PS) matching techniques [4][12]. However, no single method has gained universal acceptance due to the inherent limitations of each approach in guaranteeing comparability of the data sets.

In this paper, we introduce distance-based approaches to augment a clinical trial using partial or full ECAs. Our goal is to select a subset of RWD such that the distribution of the ECA closely matches that of the trial data. This selection process is comparable to animal studies, where subjects are assigned to treatment groups using algorithms such as Irini [5], [6], which divide the animal pool into groups with similar feature distributions. The algorithm minimizes a criterion that calculates the average of the inverse coefficients of variation across all groups and features [6]. Since Irini considers only univariate differences, a multivariate criterion was proposed [6] [13] for assigning subjects to treatment groups. This new criterion, known as the Differential *Natural Hermite Index*, is discussed below.

The proposed algorithm is an alternative to using matched pairs. The advantages over matched pairs are:

(i) The algorithm involves matching distributions and, as a result, minimizes the loss of unmatched observations. Typical matching methods use more stringent criteria, thereby resulting in data loss and reduced study power.
(ii) For smaller studies, matching could be even more difficult, resulting in additional loss of observations.

The above problems are less of an issue with distribution matching since the density estimation is less restrictive and more resistant to data loss, and generates less systematic bias.

The rest of the paper is organized as follows. In Section 2, we introduce the main idea about dissimilarity or distance indices. Section 3 describes the algorithm for data pre-processing. In Section 4, we describe the genetic algorithm and give illustrative examples involving simulations and synthetic data. In Section 5, we close with concluding remarks.

2. **Indices to measure dissimilarity among distributions**

In this section, we introduce the concept of differential projection pursuit (PP) indices and how they can be used as a dissimilarity or distance measure.

*2.1 Natural Hermite Index*

The Natural Hermite Index, introduced in Cook, Buja, and Cabrera (1993), Duan, Cabrera, Emir 2025, compares the distance between a population and the standard multivariate normal distribution. Suppose that we have a multivariate population with a cumulative distribution function (CDF) $h(x)$ on $\mathbb{R}^p$, and let $P: \mathbb{R}^p \to \mathbb{R}^d$ be a projection into the d-dimensional space with $d \ll p$, such that $f=P(h)$ is the distribution in the projection subspace $\mathbb{R}^d$. The Natural Hermite Index is defined by

$$H_N(f) = \int_{\mathbb{R}^d} \{f(x) - \phi(x)\}^2 \phi(x) dx, \quad (1)$$

where $\phi(x)$ denotes the standard multivariate normal density. The index was first proposed for application in projection pursuit, aiming to find low-dimensional projections that reveal clusters or other non-random structures in p-dimensional data. The underlying idea is that, by the Central Limit Theorem, random projections of h into $\mathbb{R}^d$ should be approximately normal, and the value of $H_N(f)$ should be small. Therefore, if a projection yields a large index value, it may be highlighting interesting, non-random structures in the original distribution, such as clusters or other nonlinear patterns.

[6] and [13] extended this concept to compare two or more distributions by introducing differential projection pursuit. When analyzing experimental data, we often deal with differential experiments, such as determining whether a new treatment is more effective than the control.

The Differential Natural Hermite Index, or Natural Hermite Index for Differential projection pursuit or dissimilarity for *k d*-

dimensional distributions $f_1(x), \ldots, f_k(x)$ is defined as follows.

Let $f_1(x), \ldots, f_k(x)$ be a set of $k$ density functions and let

$$f(x) = \frac{w_1 f_1(x) + \cdots + w_k f_k(x)}{w_1 + \cdots + w_k}$$

be the weighted average. In density estimation, the weights are functions of the sample sizes, and in some cases, when the group sizes are nearly equal, we may use $w_1 = \cdots = w_k = 1$. For every pair of densities $f_i(x), f_j(x)$, the Differential Natural Hermite Index with respect to $f(y)$ is defined by:

$$I_H(f_i, f_j) = d_f(f_i, f_j) = \left| \int_{\mathbb{R}^d} [f_i(x) - f_j(x)]^2 f(x) dx \right|^{\frac{1}{2}}. \quad (2)$$

Following [6], the Differential Natural Hermite Index may also be used as a dissimilarity measure for two or more distributions in $\mathbb{R}^d$.

The application here is the opposite of PP, which aims at maximizing the index. In our case, we want to minimize the index to search for partitions of a sample into subsamples that are identical in distribution and have differential index value near zero.

For comparison of $k>2$ populations, [6] propose minimizing:

$$C = \sum_{i<j} w_i w_j d_f^2(f_i, f_j) = \text{const} \sum_i w_i d_f^2(f_i, f)\,.$$

## 2.2 Other indices to measure distributional distances

An alternative and straightforward way to define an index for measuring the dissimilarity between two populations, based on two samples, is to use propensity scores (PS). Let X represent the set of features in the dataset, and let the binary response Y be 0 for the control sample and 1 for the new treatment sample. The PS for the $i$th observation is given by the probability $ps_i = P(Y = 1 | X = x_i)$. The values of $ps_i$ are estimated by a nonlinear function estimator, such as a Super-Learner (Van der Laan & Rose, 2011), or by a simpler logistic regression. We define the PS index as the variance of the observed propensity scores, $I_{PS}(x, y) = Var(\{ps_i\})$. This index takes the value of zero when the propensity scores function is constant.

The following result indicates that, for large sample sizes, the indices are expected to approximate zero when the samples originate from the same distribution.

**Lemma.** Given two random samples of sizes $n_1, n_2$ represented by {x=covariates, y=sample 1 or 2} from 2 distributions $f_1$ and $f_2$, and let $\hat{f}_1$ and $\hat{f}_2$ be the corresponding density estimators. If $f_1 = f_2$ then
$$\lim_{n_1, n_2 \to \infty} I_H(\hat{f}_1, \hat{f}_2) = \lim_{n_1, n_2 \to \infty} I_{PS}(x, y) = 0,$$
where n1 and n2 are sample sizes, and x and y are as defined above.

*Proof.* The consistency properties of the super-learner (Van der Laan) imply that if $n_1, n_2 \to \infty$ then $P(Y = 1 | X = x)$ converges to $n_1/(n_1 + n_2)$ for all x. Therefore, the variance $I_{PS}(x, y)$ converges to zero. The consistency of kernel density estimators (Watson, 1961) implies that $I_{PS}(\hat{f}_1, \hat{f}_2)$ also converges to zero.

Other similar indices could be derived from any method that produces an estimate of $ps_i = P(Y = 1 | X = x_i)$, such as logistic regression, LDA, deep learning, random forest, or support vector machines (SVM).

## 3. Data Pre-processing

The real-world dataset is generally less restricted than the clinical trial data, meaning that the domain of the clinical trial data is typically contained within the broader domain of the real-world data. Unlike real-world data, clinical trial subjects are selected based on specific inclusion and exclusion criteria during the screening or baseline phase. Moreover, clinical trial data may be incomplete due to participant dropouts. It has been suggested (Van der Laan (2011)) that there is, on average, a 25% dropout rate; as a result, the inclusion and exclusion criteria do not precisely define the sub-population represented by the clinical trial. Instead, this sub-population must be more accurately described by the multivariate domain of the dataset.

To select a subset of the real-world data that matches the distribution of the clinical data, all observations with any values falling outside the range of the clinical data variables are excluded. For the remaining cases, the dimension reduction algorithm described below is applied to both clinical and real-world datasets. This algorithm selects the subset of the real-world data that intersects with the convex hull of the clinical dataset after dimension reduction.

In addition to principal components, Fisher-Yates (FY) normalization (Amaratunga, et al 2014) is applied to all variables in the datasets. FY normalization converts each observation into a normal quantile based on its rank. For binary variables, FY produces values that are comparable to those generated by other transformations commonly used in deep learning, such as linear mapping of features to [0, 1] or z-score transformation. For instance, with a Bernoulli random variable where the probability of 1 is 0.01, dividing X by the standard deviation maps X=1 to 10.05, resulting in an outlier. In contrast, the FY transformation maps 1 to 2.23, with a standard deviation of 0.353.

The approach for finding the RWD inside the convex hull of the clinical study is carried out using the following algorithm:

**Algorithm 1**

(i) Apply FY to the variables.
(ii) Do a PCA and extract the first $p$ ($p \leq 7$) principal components.
(iii) Apply again FY to the principal components. Steps (i)-(iii) can be expressed as a transformation $T: \mathbb{R}^p \to \mathbb{R}^d$, where $z=T(x)$ is approximately normally distributed.
(iv) Next, the transformation $T(x)$ obtained from steps 1-3 is applied to the real-world data $z^* = T(x)$.
(v) Compute the convex hull $C_H$ of the transformed clinical study data and use it to discard the transformed real-world data that falls outside of $C_H$

Since these three steps are either linear transformations or monotonic, they are all reversible.

To illustrate the algorithm, we generated a synthetic dataset based on an abruption clinical study [17], featuring 18 predictor variables and abruption responses for about 1800 pregnancies. Using the same method with corresponding real-world data, we created another synthetic dataset representing roughly 43,000 subjects. The results are displayed in Figures 1-3.

The left panel of Figure 1 presents a scatter matrix for 6 out of the 18 variables in the clinical study dataset, with most variables being binary or categorical. The middle panel displays pairwise scatter plots of the top 5 principal components from the clinical study dataset after dimension reduction.

The right panel shows the five components of the real-world dataset following dimension reduction by Algorithm 1. Green points indicate data within the domain of the clinical study, while red points represent those outside the convex hull and are excluded from analysis. There are 15,203 subjects outside and 28,403 subjects inside the convex hull.

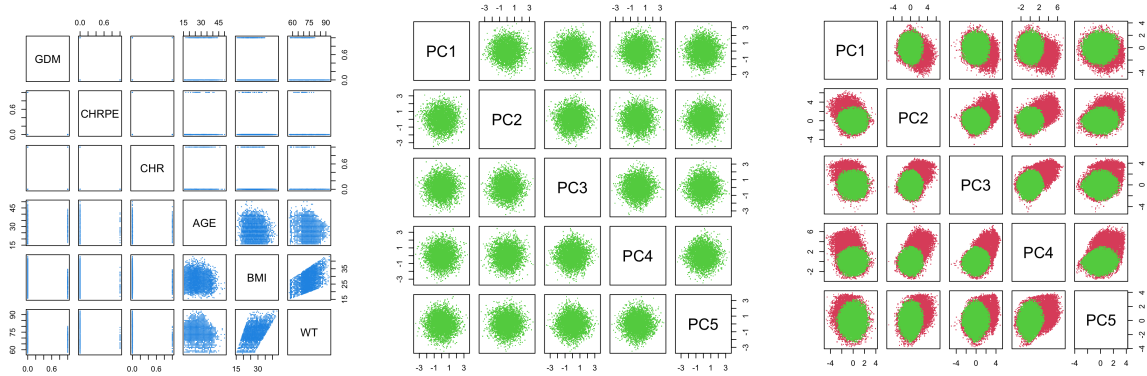

Figure 1. Illustration of dimension reduction algorithm using synthetic data

The histograms in Figure 2 demonstrate that the transformation produced by the dimension reduction algorithm results in a dataset that appears highly normally distributed. However, note that the blue histograms in the second row of Figure 2, corresponding to the transformed real-world data, reveal several differences compared to the clinical study dataset (first row). Therefore, all observations in the RWD that fall outside the convex hull of the clinical trial data should be deleted.

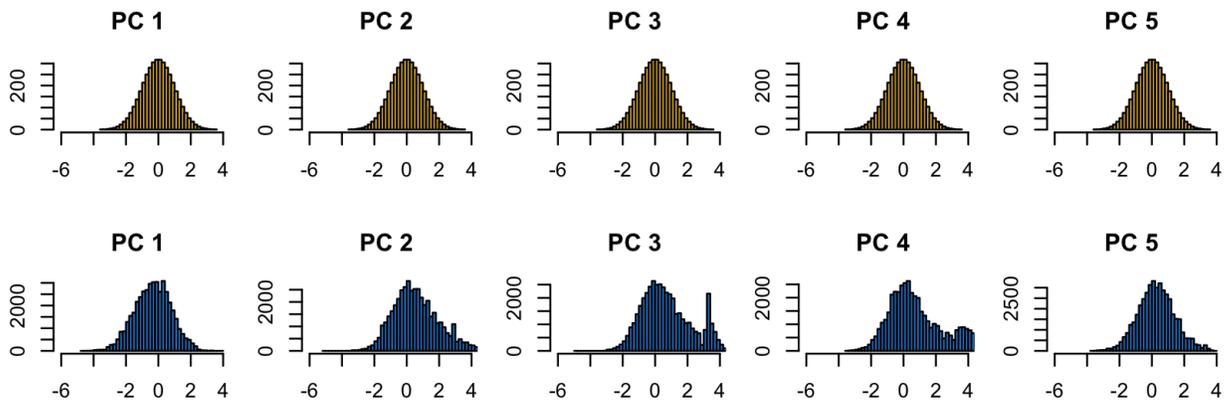

Figure 2. Histograms of transformed variables *T(x)* generated by the dimension reduction algorithm using clinical and real-world data.

To reduce computational complexity, the maximum number of principal components should be chosen based on the variance explained. Calculating the d-dimensional convex hull in step (v) of the algorithm becomes very slow for dimensions greater than seven. For d = 7, the algorithm returns a subset of 18,992 observations inside the convex hull of the clinical data, which is a significant reduction compared to the subset of 28,403 obtained with d = 5.

However, with a standard laptop, the maximum number of PCs that can be used in the algorithm is at most seven. In this example, five principal components explain 78% of the variability, whereas d = 7 explains 90% of the variability, which may be considered satisfactory. For situations where d = 7 is not satisfactory, a more efficient computational tool must be used.

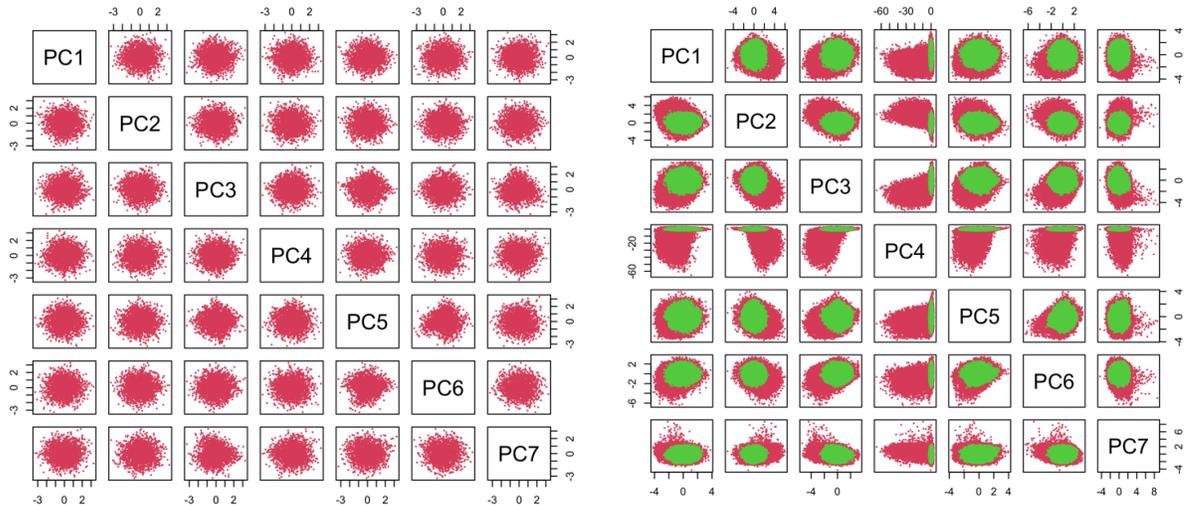

Figure 3. Plots of principal components of clinical and real-world data

In Figure 3, the Left panel shows the 7 principal components of the clinical study data after Fisher-Yates normalization. The right-side panel indicates the corresponding plots for the real-world dataset.

## 4. A Genetic algorithm for augmenting control groups from real-world data

### 4.1 Description of algorithms

In this section, we introduce a genetic algorithm based on the theory of evolution

[7][8] for augmenting a control arm of an RCT using RWD. Although Algorithm 1 selects the set of data that intersects with the convex hull of the clinical data, as shown in Figure 2, the distributions of the two data sets are not the same. The genetic algorithm is intended to correct this difference by selecting a subset of

the potential controls that match the clinical data in a multivariate distribution.

Let's assume that the data comes from a clinical study comparing a test drug (A) vs a control (B), with $n_1$ and $n_2$ subjects assigned to A and B, respectively, and $n_1 \gg n_2$. Imagine we also have a real-world dataset R of size $M \gg n_1$, which contains the same variables as those in the clinical trial dataset for A and B.

Our goal is to augment the $n_2$ subjects in the clinical trial control arm by drawing a subset B* of m observations from R. In this scenario, A represents the treatment group, comprising 4,000 subjects.

Let Y represent the outcome variable of interest, and X a d-dimensional vector of covariates. The objective is to identify B*, consisting of m observations from R, so that when combined with the $n_2$ controls in B, the distribution matches as closely as possible that of the $n_1$ observations in A, the treatment group, according to a chosen optimality criterion.

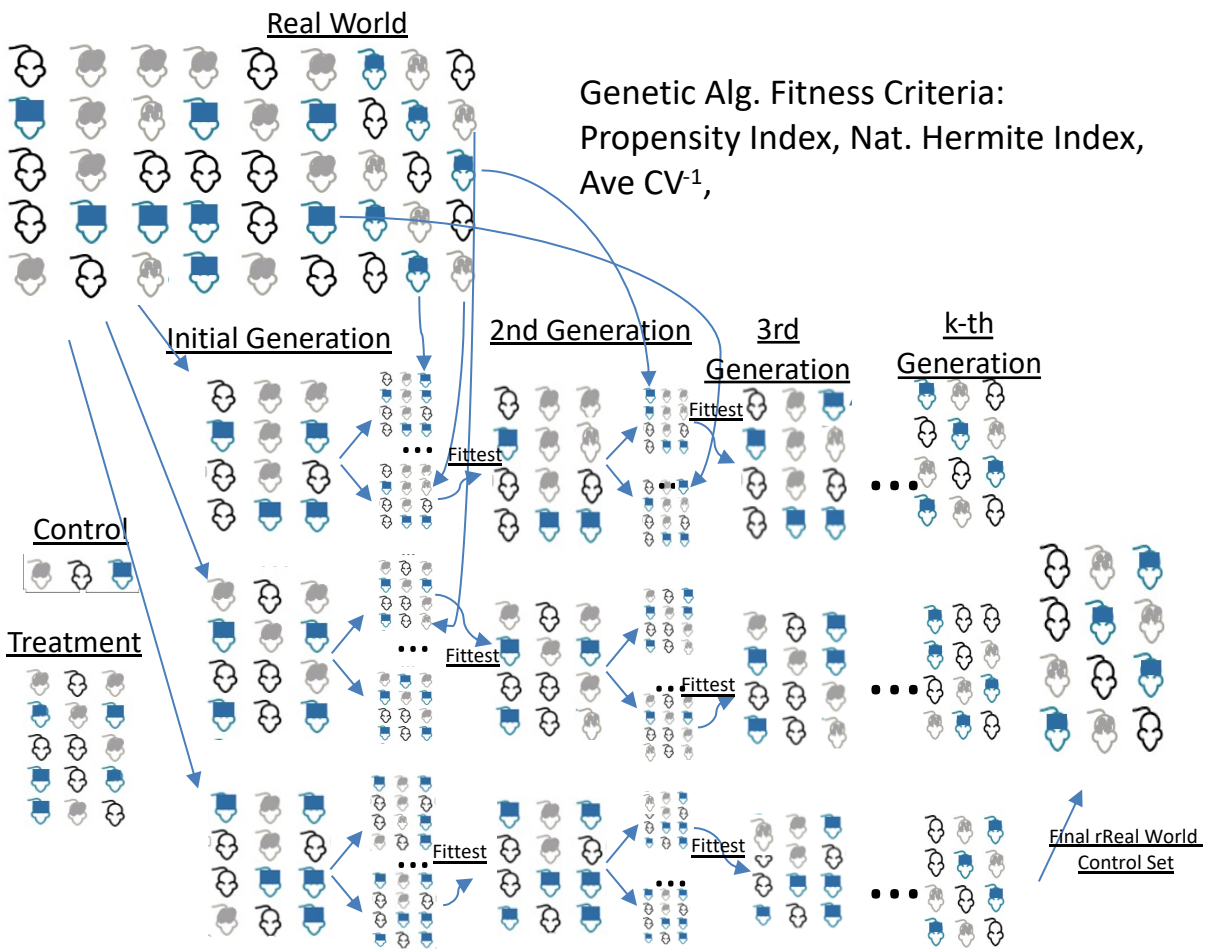

**Figure 4.** A genetic algorithm for augmenting control groups from real-world data

Figure 4 shows key aspects of a genetic algorithm. The process starts with a control set, which may be empty or not. The first generation comprises $k$ sets, created by adding real-world data to the initial control set. Each set produces $s$ mutated datasets, yielding $k \times s$ total datasets. The top $k$ sets, based on fitness, advance to the next generation. This cycle continues for $L$ generations until the convergence of the fitness criterion, which is achieved when there is no further improvement in the criterion after a few generations.

The fitness criteria here are (i) Ave $CV^{-1}$, (ii) Differential Hermite index (iii) Propensity Index described above. We will concentrate on the differential Hermite index and the propensity indices and will not consider the Ave $CV^{-1}$ because it ignores the multivariate structure of the data. In PP, we optimize and index by observing index values in a neighborhood of the current projection and choosing the next projection.

The genetic algorithm is applied to optimize the index over the space of all subsets of $R$ of size $m$.

Given the extensive number of possible subsets of $R$ of size $m$, identifying the subset that minimizes the index may not be feasible. Therefore, a genetic algorithm is employed to efficiently search for a near-optimal subset corresponding to the desired index. This algorithm generates successive generations of subsets by introducing mutations—specifically, exchanging an observation from $B^*$ with one from $R - B^*$. Each new generation is created by mutating each of the current $k$ subsets $s$ times, resulting in a larger pool of candidate subsets. Subsequently, a selection process retains only the top $k$ subsets with the best index values, ensuring the population evolves toward optimality over successive iterations.

The full algorithm consists of the following steps:

**Algorithm 2. Genetic algorithm for control set augmentation with external data.**

1. Draw an initial $k$ (e.g., $k=10$) sets of $m$ observations from R the RWD. Compute corresponding $k$ index measures $\{c_i\}$. Also calculate $C_1 = \min_{1 \leq i \leq k} c_i$

For g=2, …, G

2. Interchange one observation from each initial set of m observations by randomly drawing an observation from the RWD.

3. Compute the index measure for each augmented dataset.

4. Repeat the above $s$ times (e.g., $s=50$), forming $k \times s$ replicates of $n_2+m$ observations and $k \times s$ index measures ($c_{gij}$, $i=1, ..k$; $j=1,…,s$).

5. Select the $k$ datasets with the smallest $c_{gij}$ and compute $C_g$ the smallest $\{c_{gij}, C_{g-1}\}$

6. Repeat steps 2-5 $G$ times until $C_{g+1-\delta}$ - $C_g$ do not improve on the criterion (we used $\delta=3$)

We implemented a straightforward stopping rule in step 6: the iteration concludes when the same configuration is returned by three consecutive generations, and hence the index value remains unchanged.

When computing the propensity index using certain algorithms, such as the Super-Learner, which depends on a random k-fold cross-validation, the calculation may need to be repeated multiple times to stabilize the index value.

## 5. Simulation results

A simulation study was conducted to evaluate the impact of augmenting controls in a clinical study using a real-world data set, comparing the effectiveness of a genetic algorithm (GA) against standard randomization. This simulation was performed under the scenario where both the clinical trial data and the real-world data were generated from the same population. By construction, the GA algorithm minimizes the index that measures the difference between treatment and control populations. However, there is no guarantee that the same can be achieved through standard randomization, because the distributions of the treatment and control sets can differ by chance. Therefore, GA randomization produces better matching between the treatment and control populations, which may result in improved efficiency and greater power, as demonstrated by the simulation results below.

For this exercise, 100 datasets were generated simulating a clinical trial with a treatment group of 50 subjects and a control group of 10 subjects. The real-world database $R$ had 50,000 subjects generated following the dataset from the tutorial: **https://www.khstats.com/blog/tmle/tutorial.**

For the 50,000 subjects, the variables $W_1$, $W_2$, $W_3$, $W_4$ were generated from the following distributions:

$$W1 \sim Bernoulli(p = 0.2),$$
$$W2 \sim Bernoulli(p = 0.5),$$
$$W3 \sim Round(Unif(2,7)),$$
$$W4 \sim Round(Unif(0,4))$$

In addition, the treatment group of 50 subjects and a control group of 10 subjects were also generated from the same distributions as above. The control group of 10 subjects was augmented by selecting 40 subjects from the real-world database by one of two methods. Method 1 is standard randomization, and method 2 is using the genetic algorithm to find the subset which minimizes the propensity scores index.

Once the 100 augmentations provided by methods 1 and 2 were obtained, a response variable was generated 100 times for each dataset using the following formula:

$$Y = 5.5 + 0.2\,W_2 + log(0.1\,W_3) + 0.3\,W_4 + 0.2\,W_1\,W_4 + \delta\,Treat + \varepsilon,$$

where $\delta$ is the treatment effect size (0-5 range) and $\varepsilon$ is normally distributed with zero mean and standard deviation $\sigma_\varepsilon$ (0.1, 0.25, 0.5). Figure 5 shows power curves as a function of $\delta$, obtained from the above model using the two augmentation methods. The three panels represent the power curves corresponding to the three values of $\sigma_\varepsilon$. We conclude that in cases when the populations of treatment, control, and real-world are the same, it appears that the genetic algorithm improves power for detecting treatment effect versus control.

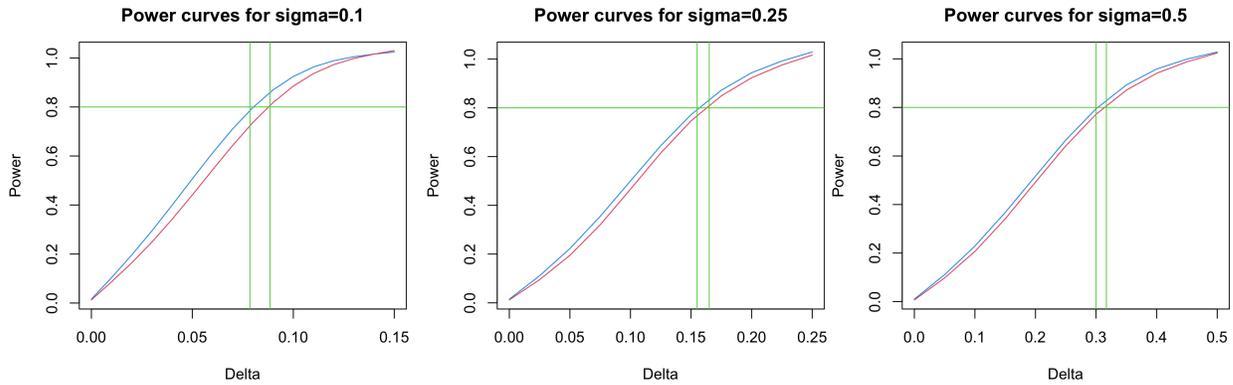

Figure 5 Simulation results comparing random augmentation (red line) to GA algorithm augmentation (blue line) for 3 values of $\sigma_\varepsilon$ (0.1, 0.25, 0.5).

### 5.2 A real-world example

This section presents the results of the genetic algorithm applied to the situation described in Section 3, where the data were pre-processed using Algorithm 1, and the subset of the RWD that intersects the domain of the treatment dataset was selected. This control subset, consisting of M = 18,992, shared the same domain as the clinical data but did not necessarily have the same distribution. Since this study aims to augment the control set to match the distribution of the treatment set, Algorithm 2 was applied after Algorithm 1. Figure 5 shows a comparison of the density estimators of the distributions of four continuous predictors in the treatment and control datasets, obtained using Algorithm 2. The density estimators are nearly identical.

Figure 6 presents a few bivariate plots of densities of the four continuous predictors, comparing the bivariate distributions in the treatment and control datasets. The results are also very similar, suggesting that the genetic algorithm achieved its purpose of balancing the treatment and controls.

The discrete/categorical variables were all binarized, resulting in 21 binary variables. After running the genetic algorithm, the proportion of 1s of each of the 21 binary variables was identical between the treatment and real-world controls, and the difference in counts of 1s was zero for all the variables.

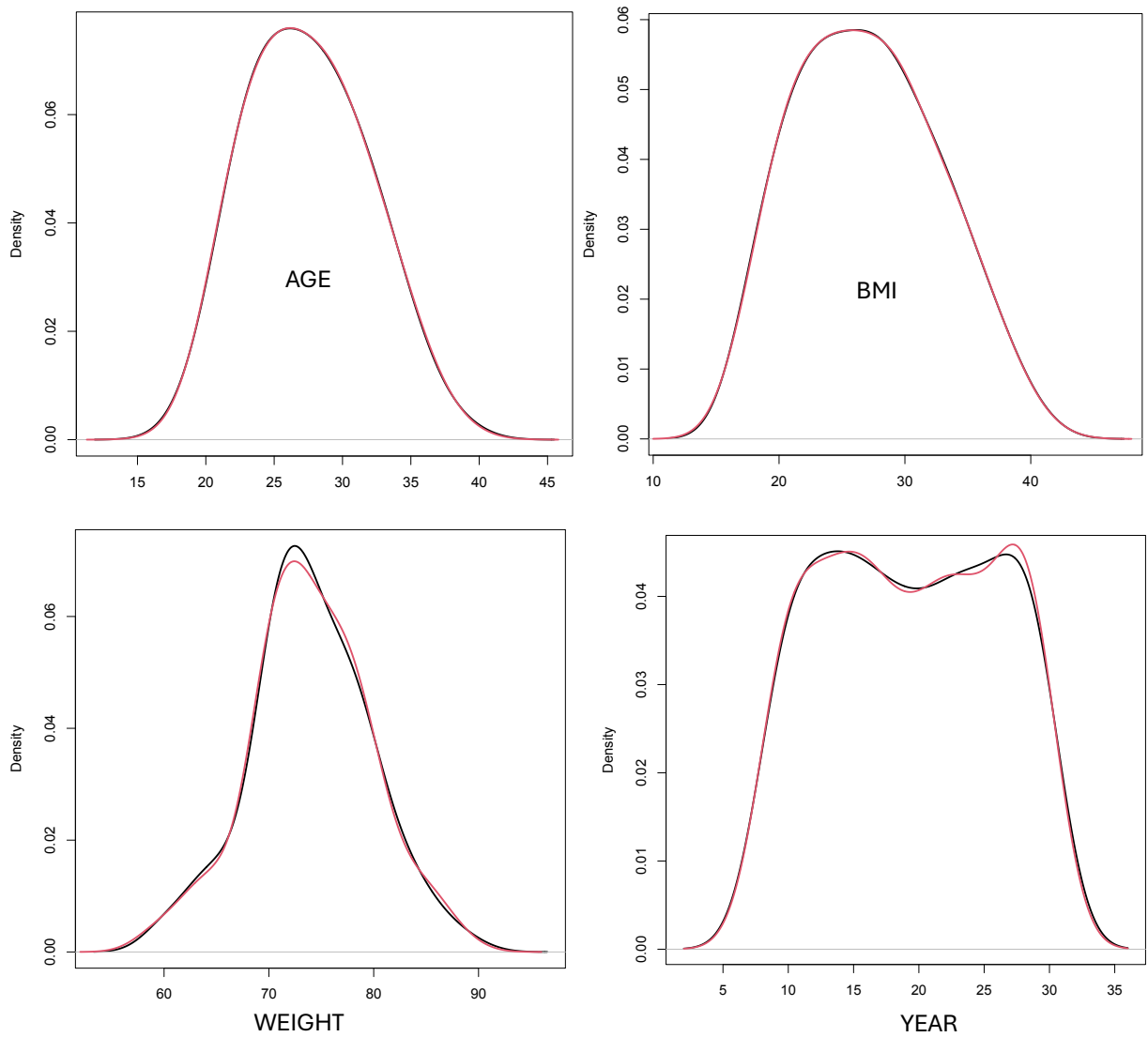

Figure 5. Density estimators of the distributions of four continuous predictors in the treatment and control datasets. The controls are a subset of the real-world data.

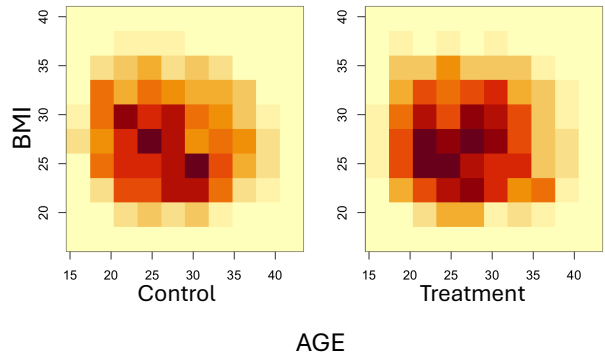

Figure 6a. Density estimators of the distributions of four continuous predictors in the treatment and control datasets. The controls dataset is a subset of the real-world dataset of controls.

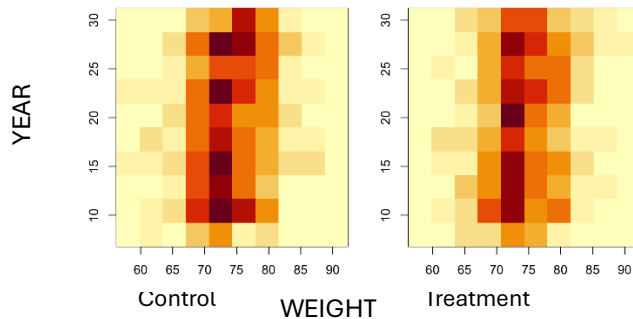

Figure 6b. Density estimators of the distributions of four continuous predictors in the treatment and control datasets. The controls dataset is a subset of the real-world dataset of controls.

## 6. Conclusion

The application of external control arms is increasingly acknowledged as a practical alternative in scenarios where randomized controlled trials are not feasible. Nevertheless, addressing challenges related to data heterogeneity is essential for achieving valid comparative analyses. Conventional matched pairs methods exhibit notable limitations, including data attrition and systematic bias arising from variations in patient health status and specificity, thereby motivating the development of alternative distribution matching methodologies. This paper introduces innovative techniques designed to enhance the reliability of augmenting control arms with real-world data. Specifically, the Natural Hermite Index and its differential projection pursuit extension are presented as multivariate criteria for quantifying dissimilarity between distributions, which facilitate the selection of subsets with aligned distributions. Additionally, an alternative measure based on the variance of estimated propensity scores is proposed to assess the similarity between treatment and control cohorts.

A preprocessing algorithm is outlined, which incorporates Fisher-Yates normalization and principal component analysis to identify and select real-world data points positioned within the convex hull of clinical trial data, thus ensuring domain compatibility. Furthermore, a genetic algorithm is utilized to iteratively refine subsets of real-world data to augment clinical trial control groups, with the objective of minimizing dissimilarity indices and optimizing distributional congruence with the treatment group.

Simulation results indicate that the genetic algorithm demonstrates substantial efficacy in identifying control subsets whose distributions closely mirror those of the treatment group, thereby strengthening the validity of subsequent statistical analyses. The combined methodological framework supports robust multivariate distribution matching, eliminating the need for matched pairs and yielding benefits in terms of reduced bias and minimized data loss.

The outcomes are promising, revealing that the treatment and control group distributions exhibit near indistinguishability. Due to computational constraints, simulations were performed on relatively small datasets; future research will encompass larger-scale simulations to further validate the effectiveness of the proposed distribution-matching methodologies.